\documentclass[prl,aps,onecolumn,showpacs]{revtex4}

\usepackage{psfrag}
\usepackage{graphicx}
\usepackage{graphics}
\usepackage{bm}
\usepackage{color}
\usepackage{centernot}

\newcommand{\be}{\begin{equation}}
\newcommand{\ee}{\end{equation}}
\newcommand{\ba}{\begin{eqnarray}}
\newcommand{\ea}{\end{eqnarray}}

\newcommand{\dcom}[1]{}
\newcommand{\dnote}[1]{}

\newcommand{\gsim}{\raise.3ex\hbox{$>$\kern-.75em\lower1ex\hbox{$\sim$}}}
\newcommand{\lsim}{\raise.3ex\hbox{$<$\kern-.75em\lower1ex\hbox{$\sim$}}}

\begin{document}

\renewcommand{\thefootnote}{\fnsymbol{footnote}}


\renewcommand{\thefootnote}{\arabic{footnote}}
\setcounter{footnote}{0} \typeout{--- Main Text Start ---}

\title{ Variable modified Newtonian Mechanics III: Turnaround radius }
\author{ C.~ C. Wong}
\affiliation{Department of Electrical and Electronic Engineering, University
of Hong Kong. H.K.}

\date{\today}
\begin{abstract}
Recently, Tension is found between Early time and Late time Hubble constant \cite{verde}-\cite{riess}. The  late time Hubble constant measurements are found to be $9\sim 10\%$ higher than the $\Lambda$CDM Hubble constant value. If these results persist, we show that it could lead to challenges to the Einstein-Straus and McVittie metrics that underlie the $\Lambda$CDM model, while a recently proposed metric (VMOND) survives.
\end{abstract}

\pacs{??}

\maketitle
\section{Introduction}
The $\Lambda$CDM model with the Planck data predicts that the Hubble constant has the present day value $H_0=67.66\pm0.42\:kms^{-1}Mpc^{-1}$ \cite{planck}. Recently,   it is reported \cite{verde}-\cite{riess} that the Hubble constant obtained from using Late time data of universe lie in the range of 
\be
H_0=73.3^{+1.7}_{-1.8} \sim 76.5\pm4.0 \:kms^{-1}Mpc^{-1},
\label{h0}
\ee
which is consistently higher than the $\Lambda$CDM prediction. An even higher value of
$H_0=82.4\pm8.4\:kms^{-1}Mpc^{-1}$  \cite{jee}-\cite{chen} is also reported using time delay measurements in strong lensing systems. There is some concensus that a tension exists between the Early time and the Late time expansion history of the universe. In \cite{verde}, it is also pointed out that the evolution of the Hubble constant offered by results from SNe and BAO do not show significant deviation from the $\Lambda$CDM model below $z\sim2$. Proposals to address the above tension highlighted in \cite{verde} mostly involve changes in Early time physics. In this work we point out that if this observational tension persists, it could also lead to violation of the turnaround radius bound derived from the metrics underlying the $\Lambda$CDM model, namely the Einstein-Straus metric and McVittie metric.

\section{turnaround radius}
The turnaround radius is the non-expanding shell furtherest away from the centre of a bound structure. Recent interest in the study of turnaround radius for structures in $\Lambda$CDM universe is started by the work in \cite{pavlidou}, followed by \cite{pavlidou1}-\cite{bhatt}.
The main argument of this work is that a primordial overdensity will eventually reach a turnaround radius $r_{ta}$, so that the outer mass shells will accelerate away, whereas the inner shells will collapse gravitationally, while the shell on the edge will move slowly to an asymptotic radius which in the case of a Einstein de-Sitter metric is given by
\begin{equation}
r_{ta} (\Lambda)= \bigg(\frac{3GM}{\Lambda}\bigg)^{1/3},
\label{ta}
\end{equation}
where $\Lambda$ is the cosmological constant in unit of $s^{-2}$. In \cite{bhatt}, it is shown that for McVittie metric, the turnaround radius has the same form as in Eq. (\ref{ta}).
In a more general setting, the authors in \cite{bhatt} consider a perturbed Friedmann-Robertson Walker metric  in Newtonian gauge 
\begin{equation}
ds^2 = -(1-2\Phi)dt^2 +(1+2\Psi)a(t)^2 \bigg( dr^2+ r^2 d\Omega\bigg)
\end{equation}
for $\Phi=\Psi=\frac{GM}{ar}$, they obtain
\be
R_{ta} =\bigg [-\frac{2GM}{(1+3\omega)H^2}\bigg] ^{1/3}=\bigg [-\frac{2GM}{(1+3\omega)H_0^2}\bigg] ^{1/3}a^{1+\omega},
\ee
where for dark energy $\omega <-1/3$ and for $\Lambda$CDM $\omega\rightarrow -1$ which gives the same result as Eq. (\ref{ta}).
\\\\
As observations from low redshift PPN parameter strongly supports Einstein Gravity \cite{collett}, we focus our attention on the solutions of Einstein Gravity for a point mass embedded in a Friedmann-Robertson-Walker (FRW) background, namely the McVittie metric, the Einstein-Straus metric and the VMOND metric \cite{wong2016}.  The VMOND metric is a solution of Einstein Gravity for a point mass embedded in a FRW background with the form
\begin{equation}
c^2ds^2 = Z c^2dt^2 - \frac{dr^2} {Z } -r^2 d\Omega^2; \:\:Z= 1-\bigg (\sqrt{\frac{2GM}{c^2r}}- \frac{H}{c} r \bigg)^2 ,
\label{s}
\end{equation}
where $H$ is the Hubble constant in units of $s^{-2}$. The energy equation for a slow speed particle is given by
\be
\dot{r}^2+\frac{h^2}{r^2}-\bigg(\sqrt{\frac{2GM}{r}}-Hr\bigg)^2 =2E_0,
\ee 
where as in Schwarzschild metric $h$ is the angular moemtnum per unit mass, and $E_0$ is the total energy. Despite the unusual appearance of the metric factor $Z$, we can see from a free falling particle without angular momentum ($h=0$) and total energy ($E_0=0$), the radial velocity has the form
\be
\dot{r}=Hr-\sqrt{\frac{2GM}{r}}
\label{dotr}
\ee
which is simply the sum of the radial velocity in a Schwarzschild spacetime and that of FRW spacetime. That is at large distances, the radial velocity is given by the FRW spacetime, and at small enough distances, the Newtonian gravitational potential dominates.
\\\\
For $H=0$ ($a = 1$) one obtains the Schwarzschild exterior metric.
For the metric factor $Z$ given in (\ref{s}), it follows that at a cut-off radius $r = r_c$ where
\be
r_c^3 = \frac{2GM}{H^2},
\label{rvmond}
\ee
one recovers a Minkowskian metric (empty space-time). This is also the place where $\dot{r}=0$ from Eq. (\ref{dotr}).
This is the radius where the metric is isotropic and homogeneous. From the Lema$\hat{i}$tre-Tolman metric in \cite{wong2016} with time $\tau$, and $r(\tau)$ as defined in \cite{wong2016}, at $r_c$, we have $dH/d\tau=0$ where the Junction conditions \cite{leibovitz}
\begin{equation}
\frac{dr(\tau)}{d\tau} =0;\;\:\: P=0
\end{equation}
is satisfied, where $P$ is the pressure which is zero at (Minkowskian) empty space.
\\\\
That is, the central point mass induces a domain of attraction of size $r_c$ inside which particles fall to the centre and beyond which they are accelerated outward by the expansion of the universe. In \cite{wong2016}, we show that the VMOND potential leads to an overdensity $\delta$ growth rate at $\delta \propto a^{2}$ (instead of the rate $\delta \propto a$ which requires additonal matter to match the observed growth rate) and VMOND potential also produces a high third acoustic peak. In \cite{wong2018}, we find that the VMOND potential effect is less dramatic once overdensity reaches order unity and the overdensity growth stops. Following theoretical and simulation work \cite{tanoglidis} and from observation \cite{martin} -\cite{molla} we conclude that there is a mass scale below which the overdensity mass shells will gravitationally collapse at high redshift, while the outer mass shell above this scale will continue to expand until it reaches its turnaround radius. The secondary infalling the mass shell (of cold hydrogen gas) will follow a dark matter potential and take a much longer time to reach the centre. However we remain agnostic in the issue of dark particle existence.
\\\\
Recall that the present value of the Hubble constant $H_0$ is given by
\be
H_0^2=\frac{8\pi G}{3} \rho_{c,0} =\frac{8\pi G}{3}\bigg( \Omega_{m,0} \rho_{c,0} +\Omega_{\Lambda,0} \rho_{c,0}\bigg) ; \:\:\:\:\: \Lambda = 8\pi G\Omega_{\Lambda,0}  \rho_{c,0} =3\Omega_{\Lambda,0}H_0^2,
\label{h0}
\ee
where $\rho_{c,0}$ is the critical density. In \cite{pavlidou}, the cosmological parameters used are $\Omega_{\Lambda,0} =0.685$ and $H_0=67.3\; kms^{-1}Mpc^{-1}$. 
\\\\
Fig.1 of \cite{pavlidou} depicts the observed turnaround radii versus the calculated bounds of structures at different mass scales. The turnaround radius can be determined observationally with a high confidence level.  The mass estimation is made by counting masses from constituent galaxies, lensing studies or X-ray spectroscopy studies. Although the counted mass can still be on a slightly lower side, a non-violation of the low mass bound would indicate a non-violation of the bound when any missing mass is included. In Fig. 1 of \cite{pavlidou}, there are structures which violate their spherical symmetric turnaround radius bound but lie within their non-Sphericity bound which is taken from simulations to be $10\%$ larger. It is clear that there are some structures such as A2055, A1918 which come very close to their non-spherical symmetric bounds. 
\\\\
The recent report by Riess et.al. \cite{riess} on the late time Hubble constant value of $H_{0,LT}=74.0\:kms^{-1}Mpc^{-1}$ has a significent implication for the turnaround radius. As described in Eq.(\ref{h0}) the value of the cosmological constant is based on $H_0$ and the density ratio $\Omega_{\Lambda,0}$.
The late time Hubble constant given in \cite{riess} is 
\be
H_{0,LT}=1.099\times H_0.
\ee
where $H_0$ is the early time ($\Lambda$CDM) Hubble constant used in the turnaround radius bound analysis in \cite{pavlidou}. As the origin of these observational discrepancies remains a topic of ongoing research \cite{verde}, we limit our attention on its implication on the turnaround radius.
\\\\
We consider firstly that there is no change to the density parameter $\Omega_{\Lambda,0}=0.685\pm 0.03$, the value of cosmological constant will increase to 
\be
\Lambda_{LT} = 1.209 \Lambda
\ee
Going back to Eq.(\ref{ta}) this new cosmological constant value will lead to a reduction of spherical symmetric turnaround radius bound to 
\be
r_{ta}(\Lambda_{LT})  =0.938\times r_{ta}(\Lambda)
\ee
and after adding non-sphericity factor it leads to the bound
\be
r_{ta, NS} (\Lambda_{LT})) =1.10\times r_{ta} (\Lambda_{LT} )=  1.032\times r_{ta} (\Lambda)
\ee 
which leads to bound violations of some observations e.g. A2055, A1918 and plausible violation of other large structures such as Virgo Cluster,  Coma Clsuter and Corona Borealis Supercluster.  Bound violation for an individual structure suggests that further studies is required. As pointed out in \cite{pavlidou} that if more accurate observations of these higher mass structures confirm that these structures are closer to the limit, or violating it, it will constitute evidence hard to reconcile with $\Lambda$CDM.  Alternatively one can consider these bound violations as an indication of a larger turnaround radius bound prediction coming from a different underlying metric. 
\\\\
If we insist that the cosmological constant remains unchanged that $\Lambda(LT)=\Lambda$, it would lead to a significant increase in the mass density parameter to $\Omega_{m,0}=0.42$, given a baryon matter density at $\Omega_b=0.05$, which would require some serious re-modelling of the matter sector of the late time universe.
\\\\
However, if the underlying metric is the VMOND metric , the turnaround radius is given by Eq.(\ref{rvmond})
\be
r_{ta}(H_{LT})  = \bigg(\frac{2GM}{H_{LT}^2}\bigg)^{1/3} =1.042\:r_{ta}(\Lambda).
\ee
Including non-sphericity factor one has
\be
r_{ta,NS} (H_{LT}) = 1.147 \:r_{ta} (\Lambda)
\ee
which remains larger than the turnaround radius bound in \cite{pavlidou}. Amongst the three metrics considered here, if the late time Hubble constant results persist, the turnaround radius bound violations will be a challenge for both the McVittie metric and the Einstein de-Sitter metric, while the VMOND metric survives. 
\\\\
Reference \cite{verde} points out that low redshift Hubble constant H(z) observations from SNe and BAO provide a guard rail for evolution of $H(z)$. In particular if we take  $H(z=0.43) =91.8\pm 5.3 kms^{-1}Mpc^{-1}$ \cite{moresco}-\cite{moresco1}, for a value of  $H_0=74.0\:kms^{-1}Mpc^{-1}$ and by using
\be
\frac{H(z)^2}{H_0^2}= \Omega_{m,0}(1+z)^3+1-\Omega_{m,0},
\ee
one would require a $\Omega_{m,0}=0.28$. For higher value of $H_0 =82.4\pm 8.4\: kms^{-1}Mpc^{-1}$ \cite{chen} it would only require a total matter density of $\Omega_{m,0} =0.124$. We notice that for a larger $H_0$ value, the matter density parameter decreases and the need for late time dark matter density also decreases. 
\\\\
In VMOND metric, for a low redshift structure, one can obtain from the VMOND a particle acceleration equation
\be
\ddot{r}=\frac{h^2}{r^3}-\frac{GM}{r^2}-H\sqrt{\frac{GM}{2r}}+\bigg(\frac{\ddot{a}}{a}\bigg) r.
\ee
An effective mass felt by the expanding background at the cut-off radius $r_c$ is
\be
M(R)= M \bigg(1+H(z)\sqrt{\frac{R^3}{2GM}}\bigg) =2M 
\ee
Within the current $H_0$ observation's upper limit value $82.4\pm 8.4\:kms^{-1}Mpc^{-1}$, it is plausible that matter density $\Omega_{b,0}=0.05$ and $\Omega_{VM} =\Omega_{b,0}$ is sufficient to explain the evolution of $H(z)$ from $z=0.43$ to present day without the need for a significant amount of dark particles at the present time. This would have implication for cosmology modelling especially for Early time. Thus more accurate data on the turnaround radius of large structures and the determination of late time Hubble constant combined could provide very useful constraints on the underlying metric for a point mass in an expanding background and the amount of non-baryonic matter density in the universe. 
\section{Summary and discussion }
Based on the $\Lambda$CDM model cosmological constant value $\Lambda$, the turnaround radius observations seem to fall inside the non-spherical symmetric bounds derived from both McVittie metric and Einstein-Straus metric. The recent observations of the late time Hubble constant $H_0$ are $9.9\%$ or higher than the $\Lambda$CDM $H_0$ value. This Hubble constant value leads to a significant reduction in the turnaround radius upper-bound of the McVittie metric and Einstein-Straus metric and certain structures in \cite{pavlidou} are found to violate the reduced turnaround radius bounds, while these data remain well inside the bound derived from the VMOND metric. Higher late time Hubble constant such as $H_0=82.4\:kms^{-1}Mpc^{-1}$ will lead to more serious violation of the turnaround radius bound for the $\Lambda$CDM model as well as a much smaller dark particle density at late time. 

\end{document}